\documentclass[prl,twocolumn]{revtex4}
\begin{document}

\title{Comment on ``Direct counterfactual transmission of a quantum state''}
\author{ L. Vaidman}
\affiliation{ Raymond and Beverly Sackler School of Physics and Astronomy\\
 Tel-Aviv University, Tel-Aviv 69978, Israel}
\begin{abstract}
 The protocol for counterfactual transmission of a qubit [Li {\it et al.}, Phys. Rev. A {\bf 92}, 052315 (2015)] relies on the counterfactuality of transmissions of bit 1 and of bit 0. Since counterfactuality of transmission of bit 0 is not established, the claim of counterfactuality of transmission of a quantum state is not founded.
\end{abstract}

\maketitle

In a paper describing direct counterfactual communication  \cite{Salih} it was claimed that a bit 1, which  Bob specifies by placing a shutter in a particular place and a bit 0, defined by placing the shutter  away from this place, are transmitted to Alice through a communication channel in a very surprising way. The result of the protocol is different evolutions of Alice's particle starting from the ready state $|R\rangle_A$ as a function of Bob's bit value:
\begin{eqnarray}\label{U1}
  |R\rangle_A |1\rangle_B &\rightarrow&  |1\rangle_A |1\rangle_B, \\
|R\rangle_A |0\rangle_B &\rightarrow&  |0\rangle_A |0\rangle_B .\label{U2}
\end{eqnarray}
 The surprising part was counterfactuality: this happened  without any particle traveling between Bob and Alice.

The linearity of quantum mechanics tells us that if the shutter is prepared in a superposition:
\begin{equation}\label{sup}
 |\psi\rangle_B=\alpha |1\rangle_B + \beta |0\rangle_B ,
\end{equation}
then the procedure  creates   entanglement between Bob's and Alice's particles  \cite{Guo}:
\begin{equation}\label{Uent}
 |R\rangle_A|\psi\rangle_B \rightarrow   \alpha |1\rangle_A |1\rangle_B  + \beta |0\rangle_A|0\rangle_B .
\end{equation}

The time symmetry of quantum processes tells us that process (\ref{Uent}) can be reversed. Bob's and Alice's particles will be disentangled and the quantum state of Bob will be restored.

The entangled state obtained after process (\ref{Uent}) is symmetric between Alice and Bob, so if in this  reversal operation the roles of Alice and Bob are switched, the state  $|\psi\rangle$ will end up in Alice's hands:
\begin{equation}\label{U-1}
 \alpha |1\rangle_A |1\rangle_B  + \beta |0\rangle_A|0\rangle_B \rightarrow   |\psi\rangle_A |R\rangle_B .
\end{equation}

Given the counterfactuality of (\ref{U1}) and (\ref{U2}), the process for creating entanglement is counterfactual and the reversed operation is counterfactual too. Thus, the two-step operation provides a counterfactual transmission of a quantum state from  Bob to Alice without particles traveling between them.

This is a gedanken experiment because we consider Alice's photon and Bob's shutter on the same footing and assume the existence of technology which can make an interference experiment with Bob's shutter. The state $|R\rangle_B $ is a quantum state of Bob's shutter exiting  Bob's interferometer toward its input port in the reversal operation. A conceptually equivalent, but a  more realistic proposal is described in  a recent paper by Li {\it et al.} \cite{Li}. (The same idea was proposed by Salih \cite{Salih2}.)

If correct, this is a far reaching result. In the words of \cite{Li}: ``An unknown quantum state can be transferred with neither quantum nor classical particle traveling in the transmission channel. We can entangle and disentangle a photon and an atom by nonlocal interaction.''

However, the issue is still under debate. Li {\it et al.} failed to mention that the counterfactuality of (\ref{U2}) was questioned \cite{Mycom,Rep,CFCF}. If (\ref{U2}) is not counterfactual, then the whole process is not counterfactual and we do not have counterfactual protocol for direct transmission of a quantum state.

The definition Li {\it et al.} adopt is: The process is counterfactual if it occurs without any real physical particle traveling between the two parties. But what is the meaning of this definition?  For the quantum particle, there is no clear definition of ``traveling''. It has a precise meaning in the framework of Bohmian mechanics, but this was not mentioned, so apparently the nonlocality of the interaction of Alice and Bob is not of the kind of the nonlocality of Bohmian mechanics. It was argued \cite{Grif} that this definition has meaning in the framework of Consisitent Histories approach \cite{Grif1}, but this approach  is  not universally agreed upon and,
 and  Li {\it et al.}  did not mention it as  well.

If the wave function of a quantum particle is not present in the transmission channel, then stating that the particle does not travel there is uncontroversial. But this is not the case in this protocol.
The statement ``the particle is not traveling in the transmission channel'' in the ``counterfactual'' protocols corresponds to the following property of the evolution of the wave function of the particle. The part of the wave function of the particle which passes through the transmission channel in its future evolution does not reach the detector detecting the particle.
 For a classical particle, whose motion is described by a trajectory, this property would be enough: if the particle traveling in the transmission channel cannot reach the detector, then the detected particle did not travel in the channel. But this classical argument cannot be applied to the past of a quantum particle \cite{past,Danan}.

In my view, the appropriate criterion for the presence of a particle in the channel is given by the analysis of the trace it leaves. If the channel is such that a single particle passing from Bob to Alice  leaves an infinitesimal trace there, then the counterfactuality of a protocol using this channel can be decided according to the trace left due to the protocol operation. The protocol cannot be named counterfactual if the trace is larger than that of a single particle. In \cite{CFCF} I have shown that this is the case for the ``counterfactual'' communication protocol \cite{Salih}. The situation for the  ``counterfactual'' transmission of a quantum state \cite{Li} is even worse. While in the ``counterfactual'' communication protocol \cite{Salih} for each bit there is a (different) part of the transmission channel without any trace, now, when we have bits in a superposition, the trace is present in all parts of the channel. Therefore, accepting the trace criterion of counterfactuality suggests that the protocol \cite{Li} should not be named ``counterfactual''.

I could not see any valid answer in \cite{Rep} to my criticism \cite{Mycom} of the counterfactuality of the protocol  \cite{Salih}
(as well as to my criticism \cite{Va07} of the counterfactuality of a similar protocol \cite{Ho}). The only known argument for the counterfactuality of these protocols relies on the notion that quantum particles move on trajectories. This notion is foreign to standard quantum mechanics.

 No paper on ``counterfactual'' communication has offered a proper definition of where a quantum particle ``travels''.
 Without such definition,  the counterfactuality  of ``Direct counterfactual transmission of a quantum state'' is not established.

This work has been supported in part by the Israel Science Foundation  Grant No. 1311/14  and the German-Israeli Foundation  Grant No. I-1275-303.14.

\end{document}